\documentclass{llncs}
\usepackage{subfigure}
\usepackage{algorithmicx}
\usepackage[ruled]{algorithm}
\usepackage{graphicx}
\usepackage{tikz}
\usepackage{amssymb}
\usetikzlibrary{chains}
\usetikzlibrary{%
		arrows,%
		shapes.misc,
		shapes.arrows,%
		chains,%
		matrix,%
		positioning,
		scopes,%
		decorations.pathmorphing,
		shadows%
	}
\usepackage{algpseudocode}

\usepackage{xxcolor}
\usetikzlibrary{chains,matrix,scopes,decorations.shapes,arrows,shapes}

\begin{document}

\title{Using Feature Models for Distributed Deployment in Extended Smart Home Architecture}

\author{Amal Tahri $ ^{1,2}$, Laurence Duchien$ ^{2}$, Jacques Pulou$ ^{1}$}

\institute{
	$ ^{1}$Orange Labs Meylan, France\\
	$ ^{2}$INRIA Lille-Nord Europe, CRISTAL laboratory,
	University Lille 1, France}
\maketitle

\begin{abstract}
Nowadays, smart home is extended beyond the house itself to encompass connected platforms on the Cloud as well as mobile personal devices. This \emph{Smart Home Extended Architecture} (SHEA) helps customers to remain in touch with their home everywhere and any time.
The endless increase of connected devices in the home and outside within the SHEA multiplies the deployment possibilities for any application. Therefore,  SHEA should be taken from now as the actual target platform for smart home application deployment. Every home is different and applications offer different services according to customer preferences.
To manage this variability, we extend the feature modeling from software product line domain with deployment constraints and we present an example of a model that could address this deployment challenge.
\end{abstract}
\section{Introduction}\label{sec:Introduction}
Smart Home Extended Architecture (SHEA) expands the Smart Home (SH) deployment environment to the Cloud and mobile personal devices to host the SH applications. 
Different domains contribute to the SHEA such as home security, comfort and energy efficiency to offer \textit{services} to customers. A service is delivered as a component-based application~\cite{Szyperski}. 
The deployment of a SH application is a mapping of a set of components onto a set of deployment \textit{nodes}. These nodes hold computational resources that must satisfy the component requirements deployed on them.\\
The variability of the SHEA comes from different points of views as it is related to the multitude of involved stakeholders for the SH market and different hardware and software resources. This variability is very challenging and has to be managed to enumerate all the deployment configurations within the SHEA. The effective deployment between the SH and the Cloud is chosen from the analysis results according to specific criteria, e.g., service availability, reduced cost.\\
Software Product Line (SPL)~\cite{Pohl2005} is a promising methodology to handle the variability.
SH applications are defined with Feature Models (FMs)~\cite{kang1990}, which are tools of SPL principles. FM is a variability modeling technique for a compact representation of all possible products, hereafter \textit{variants}, and the definition of compositional and dependency constraints, e.g., implies, excludes, among features.
Features are assets describing external properties of a product and their relationships. Constraints clarify which feature combinations are valid, named \textit{valid configurations}, using the Constraint Satisfaction Problem (CSP) solvers~\cite{apt2003principles}.
For the deployment purpose, non-functional requirements must be expressed to verify the adequacy of component requirements and node resources. FMs lack tools to express such information. Extended Feature Models (EFMs)~\cite{Czarnecki2006} overcome the FM limits by introducing non-boolean variables using \emph{attributes}, \emph{cardinalities} and \emph{complex constraints}. Attributes describe non-functional and quantified properties, e.g., CPU, RAM.
Cardinalities allow the multiplication of features and thus feature attributes such as CPU, RAM.\\
However, FM can not represent all \textit{deployment constraints}.
Deployment constraints refer to component placement indication among nodes, e.g., collocation, separation, or component requirement adequacy with the deployment node resources. EFMs are not adapted to be used in the deployment purpose as EFMs do not offer enough technical operators to express deployment constraints. Without a clear identification of deployment constraints, EFMs generate huge configuration spaces and often few \textbf{convenient} for the deployment purpose.\\
Our approach uses feature modeling to match the component requirements and the deployment node resources using CSP solver analysis. \\
The organization of the paper is the following. Motivation behind the distributed deployment across the SHEA is detailed in Section~\ref{sec:Motiv}.
The deployment oriented feature analysis is introduced in Section~\ref{DOFA}. Preliminary validation is given in Section~\ref{Primelinary-Validation}. Related work is described in Section~\ref{Related-work}. Finally, conclusion and future works are presented in Section~\ref{conclusion}.
\section{Motivations and Challenges} \label{sec:Motiv}
\subsection{Motivating Example} \label{RExample}
A customer purchases a \emph{Control Admittance} application for smart door (un)-locking based on identification mechanisms.
Different application variants are available. The basic variant represents the service of door (un)locking using the keypad identification mechanism. The person is asked to enter a pin code in the keypad to open manually the door. 
The medium variant offers the face recognition service using one recognition algorithm. When the motion detector senses a presence outdoor, the camera forwards images or video frame for face identification. This process matches the camera flow with the customer data base of authorized persons. If the person is recognized, the door opens automatically.
The premium variant offers a powerful recognition performance using multiple algorithms.
The keypad is available as a degraded mode for all variants, when Internet connection fails as it is deployed in the SH.
\begin{figure}
	\centering%
	\includegraphics[width=2in, height=1.5in]{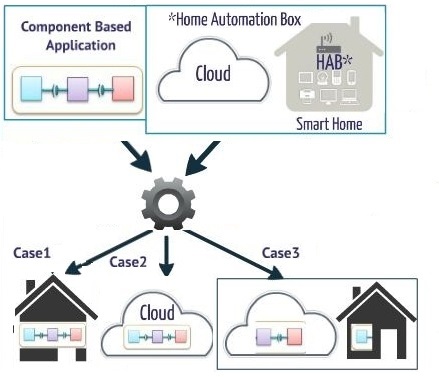}
	\caption{Deployment possibilities in the SHEA.}\label{SHEA}
\end{figure} 
Different deployment possibilities are offered to the customers between the SH and the Cloud as in Fig~\ref{SHEA}. The Home Automation Box (HAB) is an embedded environment that can host components or even a whole application. We assume that the HAB is the only node in the SH and the Cloud offers one or multiple deployment nodes, e.g., virtual machines on top of the Platform as a Service (PaaS).\\
\emph{Case 1: Deployment on the SH embedded nodes}
Application deployment in the SH confines all application components into the HAB which may lead to performance degradation because of the HAB limited resources. To satisfy component requirements, a hardware upgrade is required which raises the Bill of Material (BOM)~\footnote{http://en.wikipedia.org/wiki/Bill\_of\_materials} and, therefore, the application acquisition cost.\\
\emph{Case 2: Deployment on the Cloud} The Cloud is ``a model for enabling convenient and on-demand network access to a shared pool of configurable computing resources that can be rapidly provisioned and released with minimal management effort''~\cite{Mell2009}. The Cloud offers deployment nodes, e.g., virtual machines with on-demand resource allocation that overcome the limited capacities of the HAB.
However, the deployment on the cloud may increase the latency and response time of an application. Connection failure compromises the service availability and user experience.\\
\emph{Case 3: Deployment across the SHEA nodes}
The deployment between the HAB and the Cloud offers an attractive trade-off that overcomes the limitation presented in cases 1 and 2. As the Cloud offers on-demand resources that extend the HAB resources and reduces the application cost presented in case 1, the HAB ensures service availability when connection fails.
\begin{figure}
	\centering%
	\includegraphics[width=5in, height=1.5in]{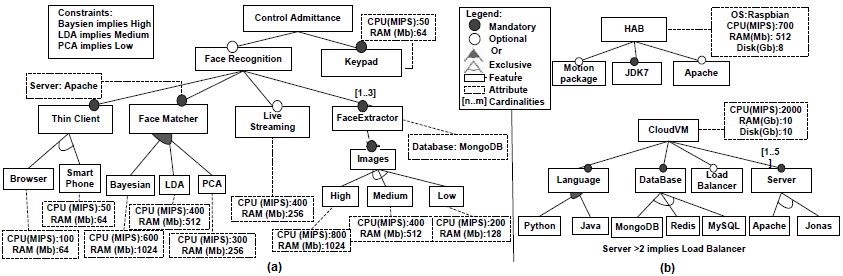}
	\caption{Control Admittance Extended Feature Model.}\label{FR1}
\end{figure} 
\subsection{Challenges}\label{subsec:challenges}
Two challenges are tackled using feature modeling in this paper:
\begin{itemize}
	\item \textit{Challenge 1 (C1)} Bridge the gap between feature modeling and deployment analysis by introducing deployment constraints in EFM.
	\item \textit{Challenge 2 (C2)} Automate the verification of deployment constraints to enumerate all the valid deployment configurations within the SHEA.
\end{itemize}
\section{Feature Analysis Oriented Deployment}\label{DOFA}
\subsection{Feature Modeling}
We refer to EFM to model the application and each deployment node. In the application EFM in Fig~\ref{FR1}(a), the components are the deployment units and represented as features. This model encompasses all the application variants presented in Section~\ref{RExample}. In Fig~\ref{FR1}(b), we present all variant for deployment nodes where features are the offered resources. The same ontology is used to declare the requirements and the resources, respectively, in the application and deployment nodes EFMs. 
\textbf{Mandatory} feature, e.g., \emph{face extractor}, represents a core functionality in the application and is always present if its parent is selected in the configuration. \textbf{Optional} feature, e.g., \emph{live streaming}, introduces the variability aspect as it may be included or not in a configuration. The \textbf{exclusive} group of the \emph{images} feature indicates that only one sub-feature can be selected in a configuration. The \textbf{or} group of the \emph{face matcher} feature allows the selection of none, one or several sub-features in the configuration. \emph{Bayesian} \textbf{implies} \emph{high} signifies that when the feature \emph{Bayesian} is selected, the feature \emph{high} must be present in this configuration. Attributes (dotted rectangles) are linked to feature to express quantified requirements, e.g., CPU and RAM. Feature cardinality (integer range [m,n], $m<=n$) determines the number of feature instances and thus the corresponding attributes allowed in the product configuration, e.g., \emph{live streaming} can be present up to three times in the same configuration. The root feature \emph{control admittance} and the \emph{keypad} feature are the basic variant. When adding \emph{face recognition} feature and choosing from the \emph{face matcher} group the \emph{PCA} feature, we obtain the medium variant.
\subsection{Approach}
\emph{Deployment Node Feature} are a new feature category representing the deployment nodes in the application EFM.  
This new category allows the separation between component features and deployment node features to declare \emph{Deployment Constraints}.
Then, we validate whether a deployment node is a suitable host for application components using feature modeling analysis.
\begin{eqnarray}
\label{eq0}
HostedBy(\mathcal{NF}, \mathcal{F}) \,
\textrm{where  $\mathcal{F} \in \mathbb{AF}$, \, $\mathcal{NF} \in \mathbb{LNF} $} \\ 
\label{eq1}
Colocated (\mathcal{F}, \mathcal{F}^{'})\,,
\label{eq2}
Separated (\mathcal{F},\mathcal{F}^{'}) \,
\textrm{where $\mathcal{F},\mathcal{F}^{'} \in \mathbb{AF}$} \\
\label{eq3}
Resource \, Constraint(r_{j}):= \sum_{l=1}^{k_{i}} R^{r_{j}}_{\mathcal{F}_{i_{l}}}  \leq  R ^{r_{j}}_{\mathcal{NF}_{i}}
\end{eqnarray}
$Hosted By$ constraint, in~(\ref{eq0}), is a binary relation between the List of deployment Node Features $\mathbb{LNF} $ and the set of Application Features $\mathbb{AF}$. The couple ($\mathcal{NF}$,  $\mathcal{F}$) implies that if the deployment node feature $\mathcal{NF}$ is selected in a configuration, then, the feature $\mathcal{F}$ is deployed on this node.\\
$Colocated$ and $Seperated$ constraints, in~(\ref{eq1}), are binary relations between two features in $\mathbb{AF}$. When both features $\mathcal{F}$ and $\mathcal{F}^{'}$ are selected in the same configuration, (i) if $Colocated$, they must be on the same deployment node. (ii) If $Separated$, they must be deployed on different deployment node. \\
$Colcated$ may be identified between (i) features of the same package that need to be deployed on the same node, (ii) features of different packages but with mutual dependencies and high coupling, (iii) all the features that contribute to the same service and should be deployed in the same network area to ensure high availability of this service in case of connection failure.\\
$Separated$ constraint can refer to (i) \emph{high availability} when two features duplicate important data that must not be lost during a single node failure, (ii) \emph{potential parallelism} when features operating independently are dispatched among different nodes to improve the throughput of the whole application, (iii) \emph{resource greedy features} when two features require a large amount of resources such as CPU or RAM, they are deployed on different nodes.\\
The $Resource \, Constraint$, in~(\ref{eq3}), ensures that the sum of the attributes for all the selected features to be hosted on embedded nodes does not exceed the node available resources. $\mathcal{F}_{i}$ is a feature to be deployed on $\mathcal{NF}_{i}$. $k_{i}$ is the size of the selected list of features on $\mathcal{NF}_{i}$ and 
$\forall r_{j}$ $\subset$ $R_{j}$,  $j$ is the resource type where $j \in[1,n]$ n being  the resource types taken into account. In our example, $n=2$ as only two resource types are considered: $r_{1} = CPU$, $r_{2} = RAM$.\\
These constraints are added in the application EFM and translated to the constraint programming Choco solver~\cite{choco} to check the configuration validity.\\
The $PossibleHost$ function below finds all $\mathcal{NF}$ that satisfy the feature attributes of $\mathbb{AF}$ and returns the analysis solution set. This function should be preceded by an initialization step that inserts the deployment constraints, i.e., $Hostedby$, $Colocated$ and $Separated$ predefined by the application developer in $\mathbb{AF}$. The algorithm takes as inputs the $\mathbb{AF}$ and the deployment node EFMs from which it constructs the list $\mathbb{LNF}$ (here $\mathbb{LNF}=\{HAB, CloudVM\}$). The algorithm has two nested $For$ loops. The outer loop covers the list $\mathbb{LNF}$. The $addFeature$ creates a mandatory feature $\mathcal{NF}$ from $\mathbb{LNF}$ under the root feature of $\mathbb{AF}$. For each node, the inner loop examines successively all the features with attributes in $\mathbb{AF}$. If no predefined $Hostedby$ constraint is found for the given $\mathcal{NF}$ and $\mathcal{F}$, the $FindMatch$ method searches the $\mathcal{NF}$ related EFM, e.g., Fig~\ref{FR1} (b), for an equivalent attribute of this $\mathcal{F}$. If match found, the $addConstraint$ method introduces a $Hostedby$ constraint to $\mathbb{AF}$ between the $\mathcal{F}$ of this attribute and the given $\mathcal{NF}$.
If no match found, the $addConstraint$ creates $not HostedBy$ constraint between the given $\mathcal{F}$ and $\mathcal{NF}$.\\
The $ResourceVerification$ procedure verifies $Resource \, Constraint$~(\ref{eq3}) and thus is only carried out for embedded nodes, e.g., in our example only HAB node is involved with the selected features hosted on them. 
The $\mathbb{AF}$ with these new constraints is translated into constraint programming and introduced to the solver (i.e., addSolver) that automatically outputs the valid configurations from where we feed the $SolutionSet \, SS$. The outer loop enters a new step and continues to scan the $\mathcal{NF}$ list until its end.\\
This section tackles the challenges in Section~\ref{subsec:challenges}.
The deployment constraints help bridge the gap between feature modeling and deployment analysis of the \textit{C1} and the algorithm automates the verification of these constraints as in \textit{C2} to enumerate the valid deployment configurations within the SHEA.
\begin{algorithm}[H]
	\caption{Matching Algorithm}\label{algo}
	\begin{algorithmic}[1]
		\State REQUIRE $Feature Model \,\mathbb{AF}$,  $list <FeatureModel> \mathbb{LNF}$
		\State ENSURE $PossibleHost$
		\Function{$ SolutionSet \, $PossibleHost}{$\mathbb{AF}, \mathbb{LNF}$}
		\State $SolutionSet \, SS = empty $
		\State Copy $\mathbb{AF}$ in $ \mathbb{AF}^{'}$
		\ForAll {$\mathcal{NF} \in \mathbb{LNF} $}
			\If {$\mathcal{NF} \notin \mathbb{AF}^{'}$} \, {$addFeature (\mathbb{AF}^{'}, \, \mathcal{NF})$}	
			\ForAll {$\mathcal{F}$ with attributes $ \in \mathbb{AF}^{'}$}
				\If {$HostedBy (\mathcal{NF}, \mathcal{F}))\notin  \mathbb{AF}^{'}$ }
					\If {$FindMatch(\mathcal{F}$ with attributes in $\mathcal{NF})$}
						\State {$addConstraint$ to $\mathbb{AF}^{'}$ ($HostedBy (\mathcal{NF}, \mathcal{F}))$}
						\State {add $\mathcal{F}$ to $\mathbb{F}_{in\mathcal{NF}}$ } \Comment{ list of $\mathcal{F}$ hosted on $\mathcal{NF}$ used in line 17 }
					\Else \, {$addConstraint$ to $\mathbb{AF}^{'}$ ($not HostedBy (\mathcal{NF}, \mathcal{F}))$}
					\EndIf
				\EndIf
			\EndFor
			\If {$ResourceVerification(\mathcal{NF},\mathbb{F}_{in\mathcal{NF} })$} 
			\Comment{check constraint~(\ref{eq3})}
			\State{$addSolver$($\mathbb{AF}^{'}$)  to $SS$}	\Comment{solver invocation}
			\EndIf
			\EndIf
		\EndFor\\
		\Return {$SS$}
		\EndFunction
	\end{algorithmic}
\end{algorithm}  
\section{Preliminary Validation} \label{Primelinary-Validation}
EFMs of the application and the deployment nodes are defined using the SALOON framework~\cite{quinton2014cloud}, for SoftwAre product Lines for clOud cOmputiNg. This framework relies on SPL principles for selecting and configuring cloud environments according to given requirements. SALOON offers the modeling and analysis tools to manage cloud variability using cardinality-based feature models and relies on the Eclipse Modeling Framework~(EMF)\footnote{http://www.eclipse.org/modeling/emf/} to present a meta-model of features. We have extended this meta-model by introducing deployment node features and deployment constraints.
We translate the features, attributes and deployment constraints to Choco solver~\cite{choco} constraint programming to check the configuration validity.
The solution evaluation computes the valid deployment configurations of the control admittance EFM on the HAB and the Cloud EFM in Fig~\ref{FR1}. Deployment constraints are introduced as follows: $HostedBy$(HAB, keypad), $Colocated$ (Baysien, live streaming), and
$Separated$ (smart phone, Baysien) and the algorithm~\ref{algo} is applied to the application EFM.
\begin{table}[h]
	\caption{Valid Configurations for Admittance Control}
	\label{validConf}
	\centering
	\begin{tabular}{|c|c|c|c|c|c|}
		\hline
		\begin{tabular}[c]{@{}c@{}}Feature \\ Model\end{tabular} & Features & Config & \begin{tabular}[c]{@{}c@{}}Config with\\1 Colocated\end{tabular} & \begin{tabular}[c]{@{}c@{}}Config with \\1 Seperated\end{tabular} & \begin{tabular}[c]{@{}c@{}}Config with\\1 Colocated\\ \&1 Seperated\end{tabular} \\ \hline
		Application & 16 & 25& 11& 16& 8 \\ \hline
		\multicolumn{1}{|l|}{\begin{tabular}[c]{@{}l@{}}Execution\\ Time (ms)\end{tabular}} & -        & 2926   & 2897 & 3639 & 2895 \\ \hline
	\end{tabular}
\end{table}
Table \ref{validConf} shows the results where the valid configuration number is reduced notably for a simple example of 16 features. In the future, realistic application set including several hundred of components will be used to characterize the limits of this method.
\section{Related Work} \label{Related-work}
The authors, in~\cite{lee2014deployment}, propose an approach for managing and verifying deployment constraints. This approach is based on Model-Driven Engineering to include deployment constraints at earlier stage of application development. The execution context includes the home devices, the mobile phones and the Cloud. The authors introduce FM to manage applications and execution context variability taking into account deployment constraints. Close to this research, our work is an extension with some differences: (i) we only consider deployment time and (ii) use CSP solver to verify the deployment configurations based on deployment constraints in feature modeling. 
Druilhe et al. in~\cite{druilhe2013energy} present a deployment model to reduce energy consumption of the home device set (Set Top Box, Gateway). They stand a distribution plan that maps the applications components on the devices considering resources and quantity of resources constraints, e.g., CPU, RAM. 
Quinton et al.~\cite{quinton2014cloud} focus on the deployment on the Cloud considering SPL techniques. They propose an extended feature models framework named SALOON to configure Cloud environments to host applications. EFM represents the Cloud environment resources, e.g., web server, data base, execution environment. This framework helps developers selecting the best solution based on specific customer criteria. We extend these previous results to include SH environment to SALOON and adapt feature modeling for the deployment purpose.
In \cite{cavalcante2012exploiting}, the authors analyze the deployment of health monitoring application variants in different Cloud platforms using SPL in order to select the possible deployment with the lower price. Our work focus on introducing deployment constraints to adapt feature modeling for deployment analysis.
\section{Conclusion}\label{conclusion}
Our approach proposes to include deployment constraint in EFM that is not proposed by other researches. 
We have used and extended the SALOON framework~\cite{quinton2014cloud} to introduce a new process for mapping application components onto deployment nodes using feature modeling. This paper raises a preliminary validation of how to adapt feature modeling for the deployment purpose. However, it does not characterize the limits of this method. The given example is restricted to SHEA with only one SH node, e.g., HAB and other examples should be checked to get better insight in the approach added value.\\

\bibliographystyle{abbrv}
\bibliography{BitexGenerale}

\begin{thebibliography}{10}

\bibitem{apt2003principles}
K.~Apt.
\newblock {\em Principles of constraint programming}.
\newblock Cambridge University Press, 2003.

\bibitem{cavalcante2012exploiting}
E.~Cavalcante, A.~Almeida, T.~Batista, N.~Cacho, F.~Lopes, F.~C. Delicato,
  T.~Sena, and P.~F. Pires.
\newblock Exploiting software product lines to develop cloud computing
  applications.
\newblock In {\em Proceedings of the 16th International Software Product Line
  Conference-Volume 2}, pages 179--187. ACM, 2012.

\bibitem{Czarnecki2006}
K.~Czarnecki, C.~Hwan, P.~Kim, and K.~Kalleberg.
\newblock Feature models are views on ontologies.
\newblock In {\em Software Product Line Conference, 2006 10th International},
  pages 41--51. IEEE, 2006.

\bibitem{druilhe2013energy}
R.~Druilhe, M.~Anne, J.~Pulou, L.~Duchien, and L.~Seinturier.
\newblock Energy-driven consolidation in digital home.
\newblock In {\em Proceedings of the 28th Annual ACM Symposium on Applied
  Computing}, pages 1157--1162. ACM, 2013.

\bibitem{choco}
N.~Jussien, G.~Rochart, and X.~Lorca.
\newblock Choco: an open source java constraint programming library.
\newblock In {\em CPAIOR'08 Workshop on Open-Source Software for Integer and
  Contraint Programming (OSSICP'08)}, pages 1--10, 2008.

\bibitem{kang1990}
K.~C. Kang, S.~G. Cohen, J.~A. Hess, W.~E. Novak, and A.~S. Peterson.
\newblock Feature-oriented domain analysis (foda) feasibility study.
\newblock Technical report, DTIC Document, 1990.

\bibitem{lee2014deployment}
K.~C.~A. Lee, M.~T. Segarra, and S.~Guelec.
\newblock A deployment-oriented development process based on context
  variability modeling.
\newblock In {\em Model-Driven Engineering and Software Development
  (MODELSWARD), 2014 2nd International Conference on}, pages 454--459. IEEE,
  2014.

\bibitem{Mell2009}
P.~Mell and T.~Grance.
\newblock The nist definition of cloud computing.
\newblock {\em National Institute of Standards and Technology}, 53(6):50, 2009.

\bibitem{Pohl2005}
K.~Pohl, G.~B{\"o}ckle, and F.~Van Der~Linden.
\newblock {\em Software product line engineering}, volume~10.
\newblock Springer, 2005.

\bibitem{quinton2014cloud}
C.~Quinton.
\newblock {\em Cloud Environment Selection and Configuration: A Software
  Product Lines-Based Approach}.
\newblock PhD thesis, Universit{\'e} Lille 1, 2014.

\bibitem{Szyperski}
C.~Szyperski.
\newblock {\em Component Software: Beyond Object-oriented Programming}.
\newblock ACM Press/Addison-Wesley Publishing Co., New York, NY, USA, 2002.

\end{thebibliography}
\end{document}